\documentclass[12pt]{article}
\usepackage{epsfig}

\begin{document}

\title{Nonlinear theory of mirror instability near threshold}
\author{E.A. Kuznetsov$^{(a,b)}$, T. Passot $^{(c)}$ and P.L. Sulem $^{(c)}$ \\
\textit{{\small $^{(a)} $ P.N. Lebedev Physical Institute RAS, 53 Leninsky Ave., 119991 Moscow, Russia}}\\
\textit{{\small $^{(b)} $L. D. Landau Institute of Theoretical Physics, 2 Kosygin str., 119334 Moscow, Russia}}\\
\textit{{\small $^{(c)} $ CNRS, Observatoire de la Cote d'Azur, PB 4229, 06304 Nice Cedex 4, France}}} 
\date{}
\maketitle

\begin{abstract}
An asymptotic model based on a reductive perturbative expansion  of
the drift kinetic and the Maxwell equations is  used to demonstrate that,
near the instability threshold, the nonlinear dynamics of mirror modes in a magnetized plasma with anisotropic ion temperatures involves a subcritical bifurcation,
leading to the formation of small-scale structures with 
amplitudes comparable with the ambient magnetic field.
\end{abstract}

\noindent
PACS: 52.35.Py, 52.25.Xz, 94.30.cj, 94.05.-a

\vspace{1cm}

{\bf 1.} In regions of planetary  magnetosheaths close to the magnetopause and
in the solar wind as well, magnetic structures with a cigar form elongated along the  direction of the ambient magnetic field  are commonly observed (see e.g. \cite{Lucek01,Sper00}). According to recent observations \cite{SLD}, more than 60\% of such structures are magnetic depressions (holes) associated with maxima of the density and pressure fluctuations.  A typical depth of  magnetic holes
is about 20\% of the mean magnetic field value and can sometimes achieve  50 \%.  The characteristic width 
of such structures is of the order of a few ion Larmor radii, and they 
display an aspect ratio of about 7-10. The origin of these structures is not
fully understood but they  are often viewed as associated with  the nonlinear development of the mirror 
instability, a kinetic instability first predicted by Vedenov and Sagdeev  \cite{VedenovSagdeev} in 1957.

The linear mirror instability has been extensively studied  both analytically
(see, e.g. \cite{PokhotelovSagdeev,H07}), and  by means
of particle-in-cell (PIC) simulations \cite{G92}.
This instability  develops in a  collisionless plasma,  when the anisotropy of the ion temperature exceeds the threshold,
\begin{equation}
{T_{\perp }}/{T_{\Vert }}-1 = {\beta ^{-1}_{\perp }}.
\label{thr}
\end{equation}
Here $\beta _{\perp }=8\pi p_{\perp }/B^2 $ (similarly,
$\beta _{\Vert }=8\pi p_{\Vert }/B^2$),
 where $p_{\perp }$ and $p_{\Vert }$ are perpendicular and parallel plasma pressures respectively.
Such conditions can be
met under the effect of the plasma compression in front of the magnetopause 
\cite{HT05}. 
As shown in  \cite{hasegawa,hall,PokhotelovSagdeev}, the instability is
arrested at large $k$   due to finite ion Larmor radius (FLR) effects.

Mirror structures are also observed
when the plasma is linearly stable \cite{EB96, Genot},  which may be viewed as 
the signature of a bistability regime. This property was also established  in the
framework of anisotropic magnetohydrodynamics, using an energetic argument \cite{PRS06}.
The aim of the present paper is to demonstrate that the bistability of mirror 
structures results from a subcritical bifurcation.
As well known, for such a bifurcation, non trivial stationary states below threshold are 
linearly unstable, while above threshold, initially small-amplitude 
solutions undergo a sharp transition to a large-amplitude state, associated with
a blowup behavior within an asymptotic theory. 
After reviewing  the  nonlinear theory of the mirror instability,
 briefly announced in 
\cite{KPS07}, we demonstrate the subcritical character of the bifurcation 
in three steps: absence of small-amplitude stationary solution above threshold,
existence of an unstable branch of solutions below threshold and blowup behavior for the
initial value problem above threshold. 

The approach is 
based on a mixed hydrodynamic-kinetic
description, assuming a weak nonlinear regime 
near threshold.  Close to  threshold, the  unstable modes have wavevectors
almost perpendicular to the ambient magnetic field $\mathbf{B}$ 
($k_{z }/k_{\perp }\ll 1$) with $k_{\perp }\rho _i\ll 1$, so that 
the perturbations can be described using a long-wave approximation.  The latter allows one 
to apply the drift kinetic equation (see, e.g., \cite
{sivukhin, kulsrud}) to estimate the main nonlinear effects
that correspond to a local shift of the instability threshold (\ref{thr}).  
All other nonlinearities connected, for
example, with ion inertia are smaller. As the result, we obtain an 
asymptotic equation for the parallel magnetic field fluctuation, $\widetilde{B_z}$ 
that displays a  quadratic nonlinearity. 
 We show that this equation
belongs to the generalized gradient type with a free energy that decreases in time, 
associated with the development of  magnetic holes. This process has a self-similar blow-up behavior. 
This means that
possible stabilization of the instability can only take place for amplitudes
of order one, a regime that is beyond the framework of the present 
asymptotics.  The present approach contrasts with the quasi-linear theory \cite{Sh64} that  
also assumes vicinity of the instability threshold but, being  based on a random phase approximation,  cannot 
predict the appearance of coherent structures. 
Phenomenological models based on the cooling of trapped particles were proposed 
to interpret the existence of deep magnetic holes \cite{KS96,Pant98}. These models
do not however address the initial value problem in the mirror unstable regime.

{\bf 2}. Consider for the sake of simplicity, a plasma with cold electrons. To describe the mirror instability in the long-wave limit it is enough to 
use the drift kinetic equation for ions ignoring parallel electric field $E_{\Vert }$
and transverse electric drift:
\begin{equation}
\frac{\partial f}{\partial t}+v_{\Vert }\mathbf{b}\cdot \nabla f-\mu \mathbf{%
b}\cdot \nabla B\frac{\partial f}{\partial v_{\Vert }}=0\mathbf{.}
\label{mainkin}
\end{equation}
In this approximation ions move along the magnetic field (${\bf b}={\bf B}/B$) due to the magnetic force  
$\mu \mathbf{
b}\cdot \nabla B$ where  $\mu =v_{\perp }^2/(2B)$ is the adiabatic invariant which  plays the role of a parameter in  equation (\ref{mainkin}). 
Both pressures $p_{\Vert }$ and $p_{\perp }$ are given by 
\begin{equation}
p_{\Vert }=mB\int v_{\Vert }^2fd\mu dv_{\Vert }d\varphi \equiv m\int
v_{\Vert }^2fd^3v,  \label{par}
\end{equation}
\begin{equation}
p_{\perp }=mB^2\int \mu fd\mu dv_{\Vert }d\varphi \equiv \frac 12m\int
v_{\perp }^2fd^3v.  \label{perp}
\end{equation}
Equation  (\ref{mainkin}) with relations (\ref{par}) and (\ref{perp}) are supplemented
with the equation expressing the 
balance of forces in a plane  transverse to the local magnetic field 
\begin{eqnarray}
&&\Pi \Big \{ -\nabla \Big( p_{\perp }+
\frac{B^2}{8\pi }\Big) \nonumber \\
&&+ \Big[ 1+\frac{4\pi }{B^2}\Big( p_{\perp}-p_{\Vert }\Big) \Big] 
\frac{\mathbf{B}\cdot \nabla \mathbf{B}}{4\pi }
\Big \}=0.  \label{balance-nl}
\end{eqnarray}
Here, consistently with the long-wave approximation, we neglect both the plasma inertia  and the non-gyrotropic contributions to the pressure tensor. Furthermore,
$\Pi _{ik}=\delta_{ik}-b_ib_k$ denotes  the  projection
operator in the plane transverse to the local magnetic field. In this equation, the first term  describes the action of the magnetic and
perpendicular pressures, the second term being responsible for magnetic lines
elasticity.

  The equation governing
the mirror dynamics is then obtained perturbatively by expanding Eqs. (\ref{mainkin}) and (\ref{balance-nl}).
 In this approach, the ion
pressure tensor elements are computed from the system (\ref{mainkin}), (\ref{balance-nl}), near a
bi-Maxwellian equilibrium state characterized by temperatures $T_{\perp
}$ and $T_{\Vert }$ and a constant ambient magnetic field  ${\bf B_0}$ taken along the $z$-direction.
 
 From Eq. (\ref{balance-nl}) linearized about the  background  field ${\bf B_0}$
by  writing 
$\mathbf{B=B}_0+\widetilde{\mathbf{B}}$  ($B_0\gg 
\widetilde{B}$) with  $\widetilde{\mathbf{B}}\sim e^{-i\omega t+i{\bf k\cdot r}}$, we have
\begin{equation}
p_{\perp }^{(1)}+\frac{B_0\widetilde{B}_z}{4\pi }=-\frac{k_{z}^2}{k_{\perp }^2}\left( 1+%
\frac{\beta _{\perp }-\beta _{\Vert }}2\right) \frac{B_0\widetilde{B}_z}{%
4\pi }.  \label{first}
\end{equation}
Here $k_z$ and $k_{\perp }$ are the projections of the wave vector ${\bf k}$, 
and $p_{\perp }^{(1)}$ is calculated from the linearized drift kinetic equation
(\ref{mainkin}):
$$
\frac{\partial f^{(1)}}{\partial t}+v_{\Vert }\frac{\partial f^{(1)}}{%
\partial z}-\mu \frac{\partial \widetilde{B}_z}{\partial z}\frac{\partial
f^{(0)}}{\partial v_{\Vert }}=0. 
$$
In Fourier space, this equation has the solution 
\begin{equation}
f^{(1)}=-\frac{\mu \widetilde{B}_z}{\omega -k_zv_{\Vert }}k_z\frac{\partial
f^{(0)}}{\partial v_{\Vert }}.  \label{omega-kin}
\end{equation}
The mirror instability is such that  $\omega /k_z\ll v_{th{\Vert }}=\sqrt{%
2T_{\Vert }/m}$. This means that the ions contributing to  the resonance $\omega
-k_zv_{\Vert}=0$, correspond to the maximum of the ion distribution function.

After substituting (\ref{omega-kin}) into the first order term for
perpendicular pressure  (\ref{perp}) and performing integration, we get
\begin{equation}
p_{\perp }^{(1)}=\beta _{\perp }\left( 1-\frac{\beta _{\perp }}{\beta
_{\Vert }}\right) \frac{B_0\widetilde{B}_z}{4\pi }-\frac {i\sqrt{\pi }\omega}
{|k_z|v_{th{\Vert }}}\frac{\beta _{\perp }^2}{\beta _{\Vert }}\frac{B_0%
\widetilde{B}_z}{4\pi }.  \label{first-perp}
\end{equation}
The first term in (\ref{first-perp}) is due to the difference between
perpendicular and parallel pressures, while the second one  accounts
for the Landau pole.

Equation (\ref{first-perp}) together with (\ref{first}) yield the
growth rate for the mirror instability in the drift approximation
where FLR corrections are neglected \cite{VedenovSagdeev} 
\begin{equation}
\gamma =|k_z|v_{th{\Vert }}\frac{\beta _{\Vert }}{\sqrt{\pi }\beta _{\perp }}
\left[ \frac{\beta _{\perp }}{\beta _{\Vert }}-1-\frac 1{\beta _{\perp }}-
\frac{k_z^2}{k_{\perp }^2\beta _{\perp }}\chi \right],   \label{growth}
\end{equation}
where $\chi= 1+(\beta _{\perp
}-\beta _{\Vert })/2$.
The
instability takes place when $\beta _{\perp }/\beta _{\Vert }-1>\beta
_{\perp }^{-1}$ and, near threshold, develops in quasi-perpendicular
directions, making the parallel magnetic perturbation dominant.

As shown in Refs. \cite{hasegawa, hall, PokhotelovSagdeev}, when the FLR 
corrections are relevant, the growth rate is modified into 
\begin{equation}
\gamma =|k_z|v_{th{\Vert}}\frac{\beta _{\Vert }\chi}{\sqrt{\pi }\beta ^2_{\perp }%
}\left[ \varepsilon-\frac{k_z^2}{k_{\perp }^2} -\frac 3{4\chi}k_{\perp }^2\rho
_i^2\right]   \label{gamma}
\end{equation}
where 
$
\varepsilon= \beta _{\perp }\chi^{-1} (\beta _{\perp }/\beta _{\Vert
}-1-\beta _{\perp }^{-1})
$
and the ion Larmor radius $\rho_i=v_{th\perp}/\omega_{ci}$ is defined 
with the transverse thermal velocity $v_{th\perp}=\sqrt{2T_{\perp}/m}$ and  the
ion gyrofrequency  $\omega_{ci} =eB_0/(mc)$. 
 This growth rate can be
recovered by expanding the general expression given in \cite{PokhotelovSagdeev}, 
in the limit of  small transverse wavenumbers. It can also be obtained  directly  from
the Vlasov-Maxwell (VM) 
equations  in a long-wave limit which retains non gyrotropic contributions \cite{CHKPST}.  
It is important to note  that the expression (\ref{gamma}) for $\gamma$ is consistent with the applicability condition $\omega
/k_z\ll v_{th{\Vert }}$, i.e. when the supercritical parameter 
$|\varepsilon| \ll 1$. In this case the instability saturation  
happens at small $k_{\perp }\propto \sqrt{\varepsilon }$ due to FLR and for almost
perpendicular direction in a small cone of angles, $k_z/$ $k_{\perp }\propto 
\sqrt{\varepsilon }$. As a result, the growth rate $\gamma \propto
\varepsilon ^2$, so that, when defining new stretched  variables by 
\begin{eqnarray}
&&k_z=\varepsilon K_z\rho _i^{-1}(2/{\sqrt 3})\chi^{1/2}, \nonumber \\
&&k_{\perp }=(2/{\sqrt 3})\sqrt{\varepsilon }K_{\perp }\rho _i^{-1}\chi ^{1/2}, \label{scaling} \\
&&\gamma =\Gamma (2/\sqrt{3})\varepsilon ^2\Omega\left( \sqrt{\pi }%
\beta _{\perp }\right) ^{-1}\left( {\chi }\beta _{\| }/\beta _{\perp
}\right) {^{3/2}}, \nonumber
\end{eqnarray}
it takes the form 
\begin{equation}
\Gamma =|K_z|\left( 1-{K_z^2}/{K_{\perp }^2}-K_{\perp }^2\right)
\label{Gamma}.
\end{equation}
Hence it is seen that, in the ($K_{\perp }-\Theta$) plane  ($\Theta\equiv K_z/K_{\perp }$),  the instability takes place inside the unit circle: 
$\Theta ^2+K_{\perp }^2<1. $ 
The maximum of $\Gamma$ is obtained for  $K_{\perp }=1/2$, $\Theta
=\pm $ $1/2$ and is equal to $\Gamma _{\max }=1/8$. Outside the circle the growth rate becomes negative (in agreement with \cite{hall}).

{\bf 3.}
As it follows from (\ref{first}), in the linear regime, near the instability threshold, the fluctuations of perpendicular and magnetic pressures 
almost compensate each other (compare with (\ref{growth})). Therefore, in  the nonlinear stage of this instability, we can expect that
the main nonlinear contributions  come from the second order corrections to the total (perpendicular plus magnetic) pressure, i.e.
\begin{equation}
p_{\perp }^{(1)}+\frac{B_0\widetilde{B}_z}{4\pi }+p_{\perp }^{(2)}+\frac{%
\widetilde{B}_z^2}{8\pi }=-\chi\frac{\partial _z^2}{\Delta _{\perp }} \frac{B_0\widetilde{B}_z}{%
4\pi }.  \label{secondorder}
\end{equation}
 This result can be obtained rigorously by means of a multi-scale expansion 
based on the linear theory scalings (\ref{scaling}). For this purpose,  we introduce a
slow time $T$ and slow coordinates ${\bf R}$ in a way 
consistent with (\ref{scaling}), and expand the magnetic field fluctuations as a powers series in $\varepsilon^{1/2}$:
\begin{equation}
\widetilde{B}_z=\varepsilon {B}_z^{(1)}+O(\varepsilon ^{2}),\,\,\,
\widetilde{\mathbf{B}}_{\perp }=\varepsilon ^{3/2}{\bf B}_{\perp }^{(3/2)}+O(\varepsilon ^{5/2}),  \label{slow-magnetic}
\end{equation}
where $ {\bf B}^{(n/2)}$ are assumed to be functions of $\mathbf{R}$ and $T$. Using these expressions it is easy to establish that quadratic nonlinear terms coming
from the expansion of $\Pi $ in (\ref{balance-nl}) as well as from the second term
in the r.h.s. of Eq. (\ref{secondorder})
are small in comparison with the  quadratic term originating from the magnetic pressure in Eq. (\ref{secondorder}).
Thus, to get a nonlinear model for mirror dynamics  it is enough to find $p_{\perp }^{(2)}$.
The expansion (\ref{slow-magnetic}) induces a corresponding expansion for the distribution function and  for both pressures. 
Defining 
$${\widetilde{p}}_{\perp }^{(n)}= \pi m \int v_\perp^2 f^{(n)}v_\perp dv_\perp dv_\|, $$
from (\ref{perp}) we have 
\[
p_{\perp }^{(2)}= (B_z^{(1)}/{B_0})^2 p_{\perp }^{(0)}+2({B_z^{(1)}
}/{B_0})\,\widetilde{p}_{\perp }^{(1)}+\widetilde{p}_{\perp }^{(2)}, 
\]
up to an additional contribution proportional to $B_z^{(2)}$ that cancels
out in the final equation due to the threshold condition.

On the considered time scale, the effect of  nonlinear Landau resonance is
negligible in the  contribution to $f^{(2)}$ that can thus be
estimated from  the equation 
$$
v_\| \frac{\partial f^{(2)}}{\partial z} +({2\mu ^2}/{v_{th\Vert }^2})%
B_z^{(1)} \frac{\partial B_z^{(1)}}{\partial z} \frac{\partial f^{(0)}}{\partial{v_\|}}  =0.  
$$
For an equilibrium bi-Maxwellian distribution, we have  
$$f^{(2)}=(2\mu^2/v_{th\Vert }^4)(B_z^{(1)})^2f^{(0)}$$
 and thus
$$
p_{\perp }^{(2)}=
\left( \beta _{\perp }-4{\beta _{\perp }^2}/{
\beta _{\Vert }}+3{\beta _{\perp }^3}/{\beta _{\Vert }^2}\right )\frac{\widetilde{B}_z^2}{8\pi}. 
$$ 
As a consequence, because of the
vicinity to threshold we obtain 
\begin{equation}
p_{\perp }^{(2)}+\frac{\widetilde{B}_z^2}{8\pi}=
\left( 1+{\beta^{-1} _{\perp }}\right)\frac{3{\widetilde{B}_z^2}}{8\pi} >0.   \label{p-perp-second}
\end{equation}
Then rewriting equation (\ref{secondorder}) using the slow variables (\ref
{scaling})  and rescaling the amplitude 
$$
{\widetilde{B}_z}/{B_0}=\varepsilon {2\chi\beta _{\perp }}(1+\beta
_{\perp })^{-1} U,
$$
we arrive at the equation  \cite{KPS07}
\begin{equation}
\frac{\partial U}{\partial T}=\widehat{K}_Z\left[ \left( \sigma -\Delta _{\perp
}^{-1}\frac{\partial ^2}{\partial Z^2}+\Delta _{\perp }^{}\right)
U-3U^2\right] .  \label{main}
\end{equation}
Here  $\sigma =\pm 1$, depending of the positive or negative sign of $\varepsilon$,
$\widehat{K}_Z=-\mathcal{H}\partial_Z$ is a positive
definite operator (whose Fourier transform is $|K_Z|$), $\widehat{H}$ is Hilbert transform 
$$
\widehat{H}f(Z)=\frac 1\pi VP\int_{-\infty }^\infty \frac{f(Z^{\prime })}{%
Z^{\prime }-Z}dZ^{\prime }.
$$
As seen from  the equation, its linear part   reproduces
 the  growth rate (\ref{Gamma}), in particular, the third term in the
r.h.s. accounts for the FLR effect.

Equation (\ref{main}) simplifies when the spatial variations are limited to
a direction  making a fixed angle with the ambient magnetic field.
After a simple rescaling, one gets 
\begin{equation}
\frac{\partial U}{\partial T} =\widehat{K}_\Xi \left[ \left( \sigma +
\frac{\partial^2}{\partial \Xi^2}\right) U-3U^2\right] ,  \label{oneD}
\end{equation}
where $\Xi $ is the coordinate along the direction of variation.
This equation can be referred to as a ``dissipative Korteveg-de Vries (KdV)
equation'', since its stationary solutions coincide with those of the  usual KdV
equation. The
presence of the Hilbert transform in Eq. (\ref{oneD}) nevertheless  leads to a
dynamics significantly different from that described by soliton equations.   
  
{\bf 4.}
Equation (\ref{main}) (and its 1D reduction (\ref{oneD}) as well) possesses 
the  remarkable property of being of the form  
$$
\frac{\partial U}{\partial T}=-\widehat{K}_z\frac{\delta F}{\delta U},
$$
where 
\begin{eqnarray}
F &=&\int \left[ -\frac \sigma 2 U^2 +\frac 12 U\Delta _{\perp }^{-1}\partial^2_ZU+\frac 12\left( \nabla _{\perp }U\right) ^2+U^3\right] d%
\mathbf{R} \nonumber  \\
\ &\equiv &- \sigma N/2+I_1/2+I_2/2+I_3 \label{free}
\end{eqnarray}
has the meaning of a free energy or a Lyapunov functional. This quantity
can  only decrease in time, since 
\begin{equation}
\frac{dF}{dt}=\int \frac{\delta F}{\delta U}\frac{\partial U}{\partial t}d%
\mathbf{R}=-\int \frac{\delta F}{\delta U}\widehat{K}_z\frac{\delta F}{%
\delta U}d\mathbf{R}\leq 0.  \label{F-time}
\end{equation}
This derivative can only vanish at the stationary localized solutions,
defined by the equation 
\begin{equation}
\frac{\delta F}{\delta U}=\left( \sigma-\Delta _{\perp }^{-1}\frac{\partial ^2}{\partial Z^2}+\Delta
_{\perp }^{}\right) U-3U^2=0.  \label{stat}
\end{equation}

We now show that  non-zero solutions  of this equation do not exist above threshold ($\sigma=+1$). For this aim, following 
Ref. \cite{kuzn-musher}, we establish relations between the integrals $N$, $I_1$, $I_2$
and $I_3$, using the fact that  solutions of Eq. (\ref{stat}) are stationary points of the functional $F$ (i.e. $\delta F=0$).
 Multiplying  Eq. (\ref{stat}) by $U$ and
integrating over $\mathbf{R}$ gives the first relation 
$$
\sigma N-I_1-I_2-3I_3=0.  
$$
Two other relations can be found if one makes the
scaling transformations, $Z\rightarrow a Z,$ $\mathbf{R}_{\perp
}\rightarrow b \mathbf{R}_{\perp }$, under which the free energy (\ref{free}) becomes a function of
two scaling parameters $a$ and $b$
$$
F(a, b)=- \frac{\sigma N}2 ab^2+ \frac{I_1}2b^4a^{-1}+\frac{I_2}2 a+I_3 ab^2. 
$$
Due to the condition $\delta F=0$, the first derivatives of $F$ at $a=b=1$ have to vanish:
\begin{eqnarray}
 \frac{\partial F}{\partial a }=-\frac{\sigma N}2-\frac{I_1}2+\frac{I_2}2+I_3 &=&0,  \nonumber \\
 \frac{\partial F}{\partial b }=-\sigma N+2I_1+2I_3 &=&0.  \nonumber
\end{eqnarray}
Hence, after simple algebra, one  gets  the three relations
\[
I_1+\frac \sigma 2N=0, \,\,\, I_3=-2I_1,\,\,\, I_2=3I_1. 
\]
For $\sigma=+1$, the first relation can be satisfied only by the trivial
solution $U=0$, because both integrals $I_1$ and $N$ are positive definite. In 
other words, above threshold,  nontrivial
stationary solutions obeying the prescribed scalings do not exist.  

In contrast, below  threshold,  stationary localized
solutions can exist. For these solutions, the free energy  is positive and 
reduces to $F_s=N/2$. Furthermore,  $I_3=\int U^3 d^3R<0$.  which means that the structures have the form of  magnetic holes.   As stationary points of the functional $F$, these solutions represent saddle points, since  
the corresponding determinant of second derivatives of $F$ with respect to scaling parameters taken at these solutions, is negative ($\partial_{aa} F\partial_{bb}F-(\partial_{ab}F)^2=-2N^2<0$). As a consequence, there exist directions in the 
eignefunction space, for which the free-energy perturbation is strictly negative,
corresponding to linear instability of the associated stationary structure.  This is one of the properties for
subcritical bifurcations. 

For the 1D model (\ref{oneD}), the  proof of instability of stationary solution $U_0=-\frac 12\mbox{sech}^2(\Xi/2)$ (which coincides with the KDV soliton) is more complicated than in 3D.  The corresponding free energy  turns out to have a minimum relatively to  the scaling parameter. Therefore, one needs to consider 
the linearized problem for perturbations $W$ $(U=U_0+W)$, which  can be formulated as 
$$
\frac{\partial W}{\partial T} =-\widehat{K}_\Xi \frac{\delta \tilde{F}}{\delta W},
$$
where 
$
\tilde{F}=\frac 12 \langle W|L|W\rangle 
$ 
is the quadratic part of the free energy and 
$$
L=1 - \frac{\partial ^2}{\partial \Xi^2}+6U_0
$$
 is the 1D Schr\"odinger operator. 

It is easily seen that the operator $L$ has one neutral (shift) mode $\psi_1=\partial_{\Xi}U_{0}$  ($L\partial_{\Xi}U_{0}=0$) associated with 
invariance by space translation, which  has one node. Thus, according the 
oscillation theorem, $L$ has one negative energy level with $E=-5/4<0$, corresponding to the ground state $\psi_0=\mbox{sech}^3(\Xi/2)$ (without nodes), which  proves the instability of the stationary solution $U_{0}$  with the growth rate equal $\frac 54 \langle \psi_0|\widehat{K}|\psi_0\rangle/\langle \psi_0|\psi_0\rangle>0$.

As a consequence, starting from general initial conditions,  the derivative $dF/dt$ (\ref{F-time}) is  almost always \textit{negative}, except for unstable stationary points (zero measure) below threshold.   In the nonlinear regime, negativeness of this
derivative implies  $\int U^3d^3R<0$, which corresponds to the formation of
magnetic holes.  Moreover, this nonlinear term (in $F$)
is responsible for collapse, i.e. formation of singularity in a finite time. 

{\bf 5.} In order to characterize the nature of the singularity of Eq. (\ref{oneD}),
it is convenient to introduce the similarity variables $\xi =
(T_0-T)^{-1/3}\Xi $, $\ \tau =-\log (T_0-T)$, and to look for a solution in the
form $U=\left( T_0-T\right) ^{-2/3}g(\xi ,\tau )$, where $g(\xi ,\tau )$
satisfies the equation 
\[
\frac{\partial g}{\partial \tau} +\frac 23g+ \frac \xi 3\frac{\partial g}{\partial \xi}
=\widehat{K}_\xi \left [ \frac{\partial^2 g}{\partial \xi^2}
-3g^2\right ] +e^{-\tau }\widehat{K}_\xi g.
\]
As time $T$ approaches $T_0$ ($\tau\to\infty$), the last term in this equation becomes negligibly small and simultaneously
$\partial_\tau g\to 0$ so that asymptotically the equation transforms into
\begin{equation}\label{asympEq}
\frac 23g+ \frac \xi 3 \frac{d g}{d \xi}
=\widehat{K}_\xi \left [ \frac {d^2g}{d\xi^2}
-3g^2\right ].
\end{equation}
For the free energy this means that close to $T_0$ the first term $\sim N$ turns out to be much smaller in comparison with all other contributions, in particular with $\int U^3 d\Xi$.

At large $|\xi |$, that corresponds to the limit $T\to T_0$, the asymptotic
solution $\widetilde{g}$ of Eq. (\ref{asympEq}) obeys 
$$
 2\widetilde{g}+\xi \frac{d \widetilde{g}}{d\xi} =C\xi ^{-2}
$$
 where $C=\frac 9\pi \int_{-\infty
}^\infty g^2(\xi ^{\prime })d\xi ^{\prime }>0$, and has the form 
$
\widetilde{g}{=C\xi ^{-2}\log \left| \xi /\xi _0\right| }$.
   For $U$, it
gives the asymptotic solution 
$$
U_{asymp}=\frac{C}{\Xi^2}\log | \Xi / \Xi_0(t)|
$$
with 
$\displaystyle{\Xi_0(t)=(T_0-T)^{1/3} \xi_0}$,
that, as $T\to T_0$, has an almost time independent tail. For 
$|\Xi |<\left( T_0-T\right) ^{1/3}|\xi_0|$ the solution is negative and becomes
singular as $\Xi $ approaches the origin.

{\bf 6.} 
Asymptotically self-similar solutions can also be constructed in three
dimensions, when rescaling the longitudinal coordinate by $(T_0-T)^{1/2}$,
the transverse ones by $(T_0-T)^{1/4}$ and the amplitude of the
solution by $(T_0-T)^{-1/2}$.
Existence of a finite time singularity for the initial value problem can be
established for initial conditions for which the functional $F$ is negative,
when the term involving $\sigma $ can be neglected, an approximation 
consistent with the dynamics:
\begin{equation}
F\rightarrow F_{\lim }\equiv \frac{I_1}2+\frac{I_2}2+I_3.  \label{F-lim}
\end{equation}
To prove this statement, consider the operator $\widehat{K}_z^{-1}$,
(inverse of the operator $\widehat{K}_z$), which is defined on functions obeying $%
\int U(Z,\mathbf{R}_{\perp })dZ=0$, a condition  consistent with Eq. (\ref{main}). 
Then the time derivative of $F_{\lim }$ can be rewritten through the
operator $\widehat{K}_z^{-1}$ as follows, 
\begin{equation}
\frac{dF_{\lim }}{dT}=-\int U_T\widehat{K}_z^{-1}U_Td\mathbf{R}\leq 0.
\label{F-reduce}
\end{equation}
Consider now the positive definite quantity 
$
\widetilde{N}=\int U\widehat{K}_z^{-1}Ud\mathbf{R}\geq 0, 
$
whose dynamics is determined by the equation 
\begin{equation}
\frac{d\widetilde{N}}{dT}=-2\left( I_1+I_2+3I_3\right) =-6F_{\lim }+I_1+I_2.
\label{N-t}
\end{equation}
Let $F_{\lim }$ be negative initially, then at $T\geq 0$ the r.h.s. of (\ref{N-t}) will be positive, 
and, as a consequence, $\widetilde{N}$ will be a growing function of time. 

Introduce  now the new quantity 
$S=-F_{lim}/ \widetilde{N}$ which is positive definite if   $F_{lim}|_{T=0}<0$.
The time derivative of $S$ is then defined by means of  Eqs. (\ref{F-reduce}) and (\ref{N-t}):
\begin{equation}
\frac{dS}{dT}=-\frac{F_{\lim }\widetilde{N}_T}{\widetilde{N}^2}+\frac 1{%
\widetilde{N}}\int U_T\widehat{K}_z^{-1}U_Td\mathbf{R}.  \label{S_t}
\end{equation}
The second term in the r.h.s. of this equation can be estimated using the Cauchy-Bunyakowsky
inequality: 
$$
\frac{d\widetilde{N}}{dT}=2\int U\widehat{K}_z^{-1}U_Td\mathbf{R}\leq 2%
\widetilde{N}^{1/2}\left( \int U_T\widehat{K}_z^{-1}U_Td\mathbf{R}\right)
^{1/2}, 
$$
that gives 
$$
\int U_T\widehat{K}_z^{-1}U_Td\mathbf{R}\geq {\widetilde{N}_T^2}/({4%
\widetilde{N}}). 
$$
Substituting the obtained estimate into Eq. (\ref{S_t}) and taking into account
definition (\ref{F-lim}) for $F_{\lim }$ and Eq. (\ref{N-t}) as well, we
arrive at the differential inequality for $S$ (compare with \cite{Turitsyn}): 
$$
\frac{dS}{dT}\geq \frac{\widetilde{N}_T}{\widetilde{N}^2}\left[ \frac{%
\widetilde{N}_T}4-F_{\lim }\right] \geq 15\,S^2.  
$$
Integrating this first-order differential inequality yields 
\begin{equation}
S\geq \frac 1{15(T_0-T)}.  \label{crit}
\end{equation}
Here the collapse time $T_0=(15\,S_0)^{-1}$ is expressed in terms of the initial value $%
S|_{t=0}=S_0$. 
It is interesting to mention that the time behavior of $S$ given by the
estimate (\ref{crit}) coincides with that given by the self-similar
asymptotics. 

{\bf 7.} In this letter, we have presented an asymptotic description of the nonlinear
dynamics of mirror modes near the instability threshold. Below threshold, we have
demonstrated the existence of unstable stationary solutions. Differently,  above threshold, no stationary solution consistent with the prescribed small-amplitude, 
long-wavelength scaling can exist. For small-amplitude initial conditions, the 
time evolution predicted by the asymptotic equation (\ref{main}) leads to a 
finite-time singularity. These properties are based on the fact that this equation   
belongs to a class of generalized gradient systems for which it is possible to
introduce a free energy or a Lyapunov functional that decreases in time.
In 1D, this model can be referred to  as the dissipative KDV
equation. The difference with the usual KDV equation is connected with the
change of the symplectic operator $\partial /\partial Z$ for the KDV case to the
positive definite operator $\widehat{K}_z=-\widehat{H} \partial /\partial Z$
for (\ref{oneD}).  This leads to significant changes in the system dynamics.

The singularity formation as well as the existence of unstable stationary structures
below the mirror instability threshold obtained with the asymptotic model,  can
be viewed as features  of a subcritical bifurcation towards a large-amplitude
state that cannot be described in the framework of the present analysis. 
Such an evolution should indeed involve saturation 
mechanisms that become relevant  when the perturbation amplitudes become
comparable with the ambient field. We note that 
recent numerical  simulations \cite{BSD03,CHKPST} of the VM equations, using either
particle-in-cell codes  or an Eulerian description, display
the formation of magnetic humps above threshold together with a phenomenon of bistability,
associated with the existence of stable large-amplitude magnetic holes both below 
and above threshold. Due to numerical constraints, these simulations
are however performed in regimes that are not sufficiently close to threshold, for the 
present theory to be applicable. 

An important question concerns the relation between the present theory of structure formation and the quasi-linear effects that could compete near threshold. An early-time quasi-linear regime could for example modify the onset of coherent structures and,
on the other hand, the development of such structures, can also affect the quasi-linear dynamics.

{\bf 8.}    This work was performed in the framework of ISSI team
``The effect of ULF turbulence and flow chaotization on plasma energy and
mass transfers at the magnetopause''. The work of EK was supported by RFBR
(grant no. 06-01-00665) and by the French Minist\`ere de l'Enseignement 
Sup\'erieur et de la Recherche during his visit at the Observatoire de la 
C\^ote d'Azur, that of TP and PLS by ``Programme National Soleil Terre'' of CNRS.

\end{document}